\documentclass[12pt]{article}

\begin{document}
\title{Two dimensional space-time symmetry in hyperbolic functions }
\author{Francesco Catoni, Paolo Zampetti
\vspace{4mm} \\ \it ENEA; Centro Ricerche Energia Casaccia;
\vspace{1mm} \\ \it Via Anguillarese, 301; 00060  S.Maria di Galeria; Roma; Italy}
\date{November 7, 2000}
\maketitle
\def\ra{\rightarrow}
\def\Ra{\Rightarrow} \def\la{\leftarrow} \def\La{\Leftarrow}
\def\lra{\leftrightarrow} \def\Lra{\Leftrightarrow}

\newcommand{\ee}{\end{equation}}    \newcommand{\be}{\begin{equation}}
\def\ba{\begin{eqnarray}}  \def\ea{\end{eqnarray}}
\def\s{\sigma} \def\S{\Sigma} \def\t{\tau} \def\r{\rho} \def\x{\xi}
\def\f{\phi} \def\c{\chi}  \def\L{\Lambda} \def\D{\Delta}
\def\z{\zeta} \def\y{\eta} \def\p{\pi} \def\o{\omega} \def\O{\Omega}
\def\a{\alpha} \def\b{\beta} \def\g{\gamma} \def\G{\Gamma} \def\d{\delta}
\def\m{\mu} \def\l{\lambda} \def\n{\nu} \def\e{\epsilon} \def\th{\theta}
\def\Th{\Theta} \def\T{\Theta}
{\bf Summary.-}
An extension of the finite and infinite Lie groups properties of complex
numbers and functions of complex variable is proposed.
This extension is performed exploiting hypercomplex number systems that
follow the elementary algebra rules. In particular the functions of such
systems satisfy a set of partial differential equations that defines an
infinite Lie group. \\
Emphasis is put on the functional transformations of a particular two
dimensional hypercomplex number system, capable of maintaining the wave
equation as invariant and then the speed of light invariant too. These 
functional transformations describe accelerated frames and can be 
considered as {\it a generalisation of two dimensional Lorentz group of special 
relativity}. As a first application the relativistic hyperbolic motion is 
obtained.  \\



\pagestyle{plain}
\section{Introduction}
The wide use of complex numbers, far beyond their algebraic introduction as
square root of negative numbers, stimulated in the past century the
search for other systems of numbers and for a generalisation of the complex
numbers properties. This led to the introduction of the so called
{\it hypercomplex numbers}, whose theory was considered completed at the beginning
of this century \cite{key3}. \\
Today hypercomplex numbers
are included as a part of abstract algebra, but a full understanding of their
implications beyond the purely mathematical ones could contribute to new
insights in many scientific fields. \\
In this paper we will use the relationship
between  the systems of hypercomplex numbers and the group theory,
pointed out by S. Lie \cite{key21} and M. G. Scheffers \cite{key5}. \\
In fact Lie showed that hypercomplex numbers define a finite group, while
Scheffers demonstrated that for functions of commutative hypercomplex
numbers the differential calculus holds as for functions of complex
variable. This implies that for functions of hypercomplex variables a system of partial
differential equations can be written  as a generalisation of the
Cauchy-Riemann conditions \cite{key9}, and that these equations define an
infinite Lie group \cite{b}.
This means that the transformations by functions of hypercomplex variables
belong to a group of the same kind of conformal group, but with specific
properties ({\it e.g.} the invariants) depending on the particular number system. \\ 
In this paper we will show that a two dimensional hypercomplex number system 
(hyperbolic numbers) is connected with  the space-time group and we will
give the physical meaning of its functional transformations, showing that
these transformations can be considered as a 
{\it generalisation of two dimensional Lorentz group of special relativity.} Moreover, we will demonstrate that the
exponential transformation (which, for functions of complex variable, gives
the equipotentials of a field produced by a point charge) yields the
constant acceleration motion in the case of hyperbolic functions.
\section{Definition of hypercomplex numbers. Some properties of the
two-dimensional systems}
The hypercomplex numbers \cite{key3,key21, a13} are defined by the
expression
\be
x=\sum^{N-1}_{\a =0} e_{\a}x^{\a}                    \label{eq:1}
\ee
where $x^{\a}$ are called {\it components} and $e_{\a}$ units or {\it versors},
as in vector algebra. The expression of eq. (\ref{eq:1}) defines a hypercomplex number if
the versors multiplication
rule is given by a linear combination of versors:
\be
e_{\a}e_{\b}=\sum^{N-1}_{\g =0} C_{\a \b}^{\g}e_{\g}    \label{eq:2}
\ee
where $C_{\a \b}^{\g}$ are real constants, called {\it structure constants}, that
define the characteristics of the system \cite{key3,key21,a13}, as we will see
in the two dimensional case considered in the following. \\ The versor product (\ref{eq:2}) defines also the product of hypercomplex numbers. This product definition makes the difference between vector algebra and 
hypercomplex systems and allows to relate the hypercomplex numbers to groups. 
In fact the vector product is not, in general, a vector while the product of
hypercomplex numbers is still a hypercomplex number; the same for the division,
that for vectors does not exist while for hypercomplex numbers, in general,
does exist. \\
Let us now consider the two dimensional hypercomplex systems and their functions
that have been recently treated in \cite{key6}. Here we will briefly recall
the main properties. In \cite{key6} the generic system of two dimensional
hypercomplex numbers is defined as:
\[w=\{ x+u\,y; \,\,\,u^2=\a+u\,\b \,\,\,\,x,\,y,\, \a,\,\b \in {\bf R}\}.\]\
For any value of $\a$ and $\b$ only three systems exist having different
properties. If we represent the structure constant $\a$ and $\b$,
in a Cartesian plane  it can be shown \cite{key6} that the parabola
$\D \equiv \b^2+4\a =0$ divides the plane in two parts, and for $\a$ and $\b$,
on the parabola or on its left and right sides we have three equivalent
systems that are called:\\ $\D <0 \,\,\,\,\,\Rightarrow$ Elliptic.\\
$\D =0 \,\,\,\,\,\Rightarrow$ Parabolic \\
$\D >0 \,\,\,\,\,\Rightarrow$ Hyperbolic \\
Numbers of the same system can be related to each other by a linear transformation,
{\it i.e.} they are equivalent. Due to this equivalence, special attention is paid
to the simplest systems ({\it Canonical systems}) \cite{key3,key21,a13,key6}
obtained with $\b=0$ and: \\  
$\a =-1\,\,\,\,\,\Rightarrow$ Elliptic, or ordinary complex numbers. \\
$\a =0 \,\,\,\,\,\Rightarrow$ Parabolic \\
$\a =1\,\,\,\,\,\Rightarrow$ Hyperbolic \\
For complex numbers is, of course, $u\equiv i$ whilst for hyperbolic numbers
we put $u\equiv h$. \\
It is known that the invariant of the group of complex numbers is
$\r^2=(x+iy)(x-iy) \equiv x^2+y^2$ {\it i.e.} the Euclidean distance.
In the same way the invariant for canonical hyperbolic system  is obtained as:
$\r_h^2=(x+h\,y)(x-h\,y) \equiv x^2-y^2$, {\it i.e.} the space-time invariant, 
that is not positive defined, and is equal to 
zero for $x=\pm y$. In the following we will call the functions of
 hyperbolic numbers {\bf hyperbolic functions}. 
The hyperbolic numbers are then related to complex numbers but,
in spite of this, they have not so widely utilised as the complex numbers
and their functions. One of the rare cases of application is reported
in \cite{key6} where the hyperbolic functions are used for the description
of supersonic effects. \\
In this paper we will point out the properties which are common to complex and
hyperbolic numbers and their differences. Moreover we will discuss the 
transformations by hyperbolic functions. In Appendix we recall some of the 
properties of the complex variable and their functions in a way that will 
allow for a direct comparison with hyperbolic variables and their functions.
\section{Hyperbolic numbers and Lorentz group}
For the hyperbolic variable one can follow an approach parallel to that
discussed in Appendix for the complex variable. Let us start pointing out 
that the Lorentz transformation corresponds to the
multiplicative group of the hyperbolic numbers. Let us write a space-time
vector as a hyperbolic variable $w=x+ht$ \cite{key6} and a
hyperbolic constant $a=a_r+ha_h$ in the exponential form\footnote{We recall 
the definition of exponential and logarithmic hyperbolic functions. The former 
is given by \cite{key6,key8}: \\$ x+ht =  \exp (X+hT)\equiv  \,\exp X \,(\cosh T+h\sinh T)$ \\
The logarithmic function can be obtained by inversion of exponential function
\cite[p. 55]{key6}:\\$ X+hT =  \ln \sqrt{x^{2}-t^{2}} + h \tanh^{-1} (t/x) $ \\
In these transformations the complete $X,\,\, T$ plane is mapped in the 
region $x>0$ and $|x|>|t|$. We note that the $x,\, \, t$ plane has the 
essential property of special relativity representative plane.\\ 
For the constant $a=a_r+ha_h$ in exponential form we assume $|a_r|>|a_h|$.}: \\
$a_r+ha_h \equiv \exp (\r_h \,+h\th_h)\equiv \exp \r_h \, (\cosh \th_h +
h\sinh \th_h ) \mbox{ where }
\r_h=\ln \sqrt {(a_r^2-a_h^2)} ; \,\,\,\th_h =\tanh^{-1} (a_h/a_r).$ \\
Then the multiplicative group, $w^{\prime}\equiv\ x^{\prime}+ ht^{\prime}=a\,w$
becomes:
\begin{equation}
\label{eq:8} x^{\prime}+ ht^{\prime}=\sqrt {(a_r^2-a_h^2)}\left[x\,
\cosh \th_h +t\,\sinh \th_h +h(x\,\sinh \th_h +t\,\cosh \th_h )\right]
\end{equation}
In this equation by letting $(a_r^2-a_h^2)=1$ and considering as equal
the coefficients of the versors ``1'' and ``h'', as we make in
complex analysis, we get the Lorentz transformation of special 
relativity \cite{ug}. It is interesting to note that the same result is
normally achieved by following a number of ``formal" steps \cite[p.94]{key4},
\cite[p.50]{key10},
{\it i.e.} by introducing an ``imaginary" time $t^{\prime}=it$ which makes the
Lorentz invariant ($x^2-t^2$) equivalent to the rotation invariant ($x^2+y^2$),
and by introducing the hyperbolic functions through their equivalence with
circular functions of an imaginary angle. Let us stress that this procedure
is essentially formal, while the approach based on hyperbolic numbers leads to
{\it a direct description of the Lorentz transformation of special
relativity explainable as a result of symmetry (or invariants)
preservation}: the Lorentz invariant (space-time ``distance'') is the 
invariant of hyperbolic numbers. 
Therefore we can say that the hyperbolic numbers have the {\it space-time 
symmetry}, while the complex numbers have the symmetry of two spatial 
variables, represented in a Euclidean plane. Within the limits of our 
knowledge, the first algebraic description of Special Relativity, directly
by these numbers (called ``Perplex numbers"), has been introduced in
\cite{key8}.
With the exposed formalism we can see that the Lorentz transformation is 
equivalent to a ``hyperbolic rotation''.
In fact let us write in the Lorentz transformation, the hyperbolic variable 
$x+ht$ in exponential form:\\
$x+ht=\exp \r(\cosh\th+h \sinh \th) \mbox{ where }
\r=\ln\sqrt{(x^2-t^2)} \,\,\,,\,\,\,\th=\tanh^{-1}(t/x).$  Then:
\be 
w^\prime =aw=\exp\r[\cosh(\th_h+\th) +h\sinh(\th_h+\th)] \label{rh}
\ee
From this expression we see that the Lorentz transformation is equivalent 
to a ``hyperbolic rotation'' of the $x+ht$ variable. Then the invariance
under Lorentz transformation can be also expressed as an independence on
the hyperbolic angle $\th$.

\section{Physical meaning of transformations by hyperbolic functions}
Let us now consider the functions of hyperbolic variable. As for the functions
of complex variable, the functions $f(w)=u(x,t)+h v(x,t) $ are said to be
functions of the hyperbolic variable $w=x+ht$ and consequently they are
called {\it hyperbolic functions} if $u,\, v$ satisfy the following system
of partial differential equations \cite{key6}:
\begin{equation}
\label{CRL}u_{,x}=v_{,t} \, ;\, \, \, u_{,t}=v_{,x}
\end{equation}
where the comma stands for derivation with respect to the variables which follow.
From these equations, it follows that the relationship between the hyperbolic
functions and the wave equation is the same as that existing between the
functions of complex variable and the Laplace equation \cite{key9,key6}.
Accordingly, the wave equation is satisfied  by hyperbolic functions and it
is invariant for the transformations $(x,t)\!\Rightarrow\!(u,v) ,$ with
$u, \, v$ satisfying the system of eq.s \ref{CRL}. Therefore $t$ can
be referred to as a physical normalized (speed of light c=1) time variable. 
The invariance of wave equation means that  the speed of light does 
not change if the coordinate system is changed by these functional 
transformations, {\it i.e.} that the well known postulate of special 
relativity is also valid for all these coordinate
systems. Consequently this infinite group of functional transformations can
be considered as a {\it generalisation of Lorentz-Poincar\'e two dimensional 
group.} \\ Now let us emphasize the physical meaning of the
functional transformations. It is known that the linear Lorentz
transformation of special relativity represents a change of inertial
frame. Relating  the space and time variables to Cartesian coordinates, the
inertial motion is represented by straight lines and, applying the linear
Lorentz transformation, these lines remain straight lines. The functional
transformations change the straight lines in one reference frame into
curved lines in the other frame. From a physical point of view, a curved
line represents a non inertial motion, {\it i.e.} a motion in a field. The
question which arises is whether these functional transformations represent
any physical fields. Now we show that extending to hyperbolic functions the 
procedure reported in Appendix for complex variable,  we obtain the 
relativistic hyperbolic motion. \\
For the $x,t$ variables the experimental evidence for symmetry is provided by 
the invariance under the Lorentz transformation, or by eq. \ref{rh},
the independence on the hyperbolic angle $\th$. Thus let us find a solution 
$U(x,t)$ of the wave equation $U_{,xx}-U_{,tt}=0 $ independent of the 
hyperbolic angle. We can proceed as in Appendix, and write the hyperbolic 
variable $x+ht$ as exponential function of the variable $X+hT$.
Then with the transformation: \\$
x=\exp X\,\cosh T,\,\,t=\exp X\,\sinh T$
or $X= \ln \sqrt{x^2-t^2},\,\, T=\tanh {}^{-1}(t/x)$
we must solve the equation:
\begin{equation}
U_{,XX}-U_{,TT}=0 \label{onde}
\end{equation}
The $U$ invariance for ``hyperbolic rotation'' means independence on $T$ 
variable. Therefore $U_{,TT}\equiv 0$, and $U$ will depend
only on the variable $X$. The partial differential equation, eq. \ref{onde},
becomes a normal differential equation $d^2U/dX^2=0$, with the elementary
solution: 
\begin{equation}
\label{eq:11}U=AX+B \equiv A \ln \sqrt{(x^{2}-t^{2})} +B
\end{equation}
If we consider, as for the Laplace equation, the ``equipotentials''
$U=$const,
in the $X,\, T$ plane these lines are straight lines, that can be expressed as
$X=$const$\equiv \ln g,\,\, T=g^{-1}\tau$. Going over 
to $x,\,t$ variables these straight lines become
the hyperbolas: $$x=g \, \cosh g^{-1}\tau \, ; \, \, t= g \, \sinh
g^{-1}\tau $$ that represent the hyperbolic motion \cite[p.166]{key10},
where $\t$,  the time coordinate in a reference frame in which $X=$const,
is the proper time.  This is the law of motion of a body in the field of a
constant force, calculated by the relativistic Newton's dynamic law. In
this example the motion has  been obtained as a transformation  through the
exponential function of a particular straight line. This transformation
belongs to a
group that preserves the symmetry in space-time, and, in so doing, it
satisfies widely accepted concepts of modern physics \cite[p.48]{key11}.
\section{Conclusions}
In summary, we can conclude that:
the space-time symmetry is best treated by means of hyperbolic numbers 
and the space-time field is best described by functions of hyperbolic
variable, following the fundamental principle according which a significant
simplification in any physical problem can be obtained by using a
mathematical approach with the same symmetry of the problem. \\
\appendix 

\section{Group properties of complex numbers and of functions of complex
variable}
The complex numbers can represent plane vectors and the related linear algebra
\cite[p.73]{key4}: $z=x+iy$ is interpreted as a vector of components $x$
and $y$, and versors $1$ and $i$. In vector form we write $\vec z=\vec
1x+\vec {\i}y$, where $x$ and $y$ are the current coordinates of the plane.
If we consider the multiplication by a constant:
\be
z_1=az\equiv
(a_r+ia_i) (x+iy),        \label{1}
\ee
the complex numbers play the role of both vector and operator (matrix)
\cite[p.73]{key4} and eq. (\ref{1}) is equivalent to the familiar expression
of linear algebra: $$
\left(
\begin{array}{c}
x_1 \\
y_1
\end{array}
\right) =\left(
\begin{array}{cr}
a_r & -a_i \\
a_i & a_r
\end{array}
\right) \left(
\begin{array}{c}
x \\
y
\end{array}
\right).
$$
If we write the constant $a$ in the exponential form: \\ $a\equiv
(a_r+ia_i)=\exp \r \,(\cos \f +i\sin \f)$, where  $\r=\ln \sqrt{a_r^2+a_i^2}
$, $\f =\tan {}^{-1}a_i/a_r$, \\ we see that the constant $a$ plays
the role of an operator representing an orthogonal axis rotation with a
homogeneous dilatation (homothety). If $\r=0$, and if we add another
constant $b=b_r+ib_i$, then $z_1=az+b$ gives the permissible vector transformations in
a Cartesian plane. In group language this is the Euclidean group of
roto-translations which depends on the three parameter  $\f,\,\,b_r,\,\,b_i$.
Then we can use complex numbers to describe plane vector algebra
because the vectors are, usually, represented in an orthogonal coordinate
system and the additive and multiplicative groups of complex numbers are
related to Euclidean group.
In mathematical physics, even more important is the conformal group deriving
from functions of the complex variable. \\ According to Riemann \cite{key9}: a
function $w(z)=u(x,y)+iv(x,y)$ is said to be a function of the complex
variable $z$ if its derivative is independent of direction. If this
condition is verified, the partial derivatives of $u,\,v$, satisfy the
Cauchy-Riemann's (C-R) equations:
\be
u_{,x}=v_{,y} \, ;\, \, \,u_{,y}=-v_{,x} \label{2}
\ee
where the comma stands for derivation with respect to the variables which follow.
Thanks to eq. (\ref{2}), $u$ and $v$ satisfy the Laplace equation $U_{,xx}+U_{,yy}=0$,
which is invariant under the transformations $x,\,\,y\Rightarrow u,\,\,v$,
where $u,\,\,v$ are given by the real
and imaginary part of the same function of complex variable \cite{key9}.
In the language of the classical Lie groups \cite{key21,b}, the Euclidean 
group is said to be a finite group
because it depends on three parameters, whilst the conformal group is said
to be infinite because it depends on arbitrary functions. The Euclidean
group is very important in itself. Moreover, if considered as addition and
multiplication of complex constants and variables, it represents the simplest
subgroup of the conformal group. Both groups derive from the symmetries related to the
``operator" ``i"  of the complex variable. \\ A connection between these two
groups can be found if one looks for a field that satisfies the Laplace
equation and that is invariant for the rotation group. Having set these
requirements the problem is equivalent to calculate the potential of
a central field, that is a function only of the source distance.
This problem is usually solved \cite[p.341]{key9} by means of a polar
coordinate transformation ($x,\,y \Rightarrow \r,\,\f$), which transform
the Laplace equation in: $u_{,\r \r}+\r^{-1}u_{,\r}+\r^{-2}
u_{,\f \f}=0.$ \\ The use of a complex exponential transformation, that
has the same symmetries of the polar one, leaves the Laplace equation
invariant; then with the transformation: \\$
x=\exp X\,\cos Y,\,\,y=\exp X\,\sin Y$
or $X= \ln \sqrt{x^2+y^2},\,\, Y=\tan {}^{-1}(y/x)$
we must solve the equation:
\begin{equation}
U_{,XX}+U_{,YY}=0 \label{eq:10}
\end{equation}
The $U$ invariance for rotation means independence on the rotation angle,
represented here by $Y$. Therefore $U_{,YY}\equiv 0$, and $U$ will depend
only on the variable $X$. The partial differential equation, eq. (\ref{eq:10}),
becomes a normal differential equation $d^2U/dX^2=0$, with the elementary
solution: $$U=aX+b\equiv a\ln \sqrt{x^2+y^2}+b$$ which represents
the potential of a point charge. Then in the $X\,,\, Y$ plane the straight
lines $X=const.$ give the equipotential, and, as is better known, the
circle $x^2+y^2= const.$ are the equipotential in the $x\,,\, y$
plane. We can note that from the ``symmetry"
of the finite rotation group, we have obtained the Green function for the
partial differential Laplace equation.

\end{document}